\documentclass[aip,twocolumn]{revtex4}

\usepackage{amssymb}
\usepackage{amsmath}

\def\ov#1{\overline{#1}}

\def\vb#1{\mbox{\boldmath$#1$}}
\def\pd#1#2{\frac{\partial #1}{\partial #2}}

\def\wh#1{\widehat{#1}}
\def\bdot{\,\vb{\cdot}\,}
\def\btimes{\,\vb{\times}\,}

\def\bhat{\wh{{\sf b}}}
\def\exd{{\sf d}}

\newcommand{\bc}{\begin{center}}
\newcommand{\ec}{\end{center}}
\newcommand{\bt}{\begin{tabbing}}
\newcommand{\et}{\end{tabbing}} 
\newcommand{\be}{\begin{eqnarray*}}
\newcommand{\ee}{\end{eqnarray*}}

\begin{document}

\title{Equivalent representations of higher-order Hamiltonian guiding-center theory}

\author{Alain J.~Brizard$^{1,2}$ and Natalia Tronko$^{2,3}$}
\affiliation{$^{1}$Department of Physics, Saint Michael's College, Colchester, VT 05439, USA \\ 
$^{2}$Centre de Physique Th\'{e}orique, Campus de Luminy, Case 907, 13288 Marseille cedex 9, France \\ 
$^{3}$Centre for Fusion, Space and Astrophysics, Department of Physics, University of Warwick, Coventry CV4 7AL, UK}

\begin{abstract}
Two complementary representations of higher-order guiding-center theory are presented, which are distinguished by whether higher-order corrections due to magnetic-field nonuniformity appear in the guiding-center Poisson bracket or the guiding-center Hamiltonian. The equivalence of these two representations implies that the guiding-center gyroaction adiabatic invariant and the guiding-center Hamilton equations of motion are representation-invariants. By using a new perturbative hierarchy in guiding-center theory, second-order corrections are introduced in Hamiltonian guiding-center theory without the need to calculate the guiding-center transformation at second order.
\end{abstract}

\begin{flushright}
May 24, 2012
\end{flushright}

\maketitle

Guiding-center theory \cite{Northrop_63,CB_09} is a powerful paradigm for the reduced description of charged-particle dynamics in weakly-nonuniform magnetic fields, which yields a deep understanding of the physics of magnetic confinement in laboratory and space plasmas over long time scales 
\cite{White_Chance}. In his pioneering work, Littlejohn \cite{RGL_83} used Lie-transform perturbation methods \cite{RGL_82} to derive noncanonical Hamiltonian guiding-center equations of motion that contain first-order corrections associated with the magnetic-field nonuniformity. These first-order corrections not only have theoretical importance in constructing a hierarchy of adiabatic invariants (e.g., \cite{Northrop_66}) but also play a significant role in ensuring the accuracy of guiding-center numerical codes \cite{Belova_03} in following magnetically-trapped particles in complex magnetic fields.

The first attempts at keeping second-order corrections in the expressions for the adiabatic invariants appeared in non-Hamiltonian guiding-center equations \cite{HTH,Northrop_Rome}. Hamiltonian guiding-center equations that include second-order corrections (appearing in the guiding-center Hamiltonian) were recently derived by Parra and Calvo \cite{Parra_Calvo}, who pointed out that these higher-order corrections were required in order to construct a consistent Hamiltonian gyrocenter theory \cite{Brizard_Hahm} that retains all second-order corrections due to magnetic nonuniformity as well as fluctuation amplitude and space-time scales. Because of their potential relevance in numerical gyrokinetic applications \cite{gyro}, a complementary higher-order Hamiltonian guiding-center theory is constructed here, which includes second-order corrections (appearing in the guiding-center Poisson bracket), by using a new perturbative hierarchy in guiding-center theory.

Modern guiding-center Hamiltonian theory \cite{CB_09} is described in terms of noncanonical guiding-center phase-space coordinates $z^{\alpha} \equiv 
({\bf X},p_{\|},\mu,\theta)$, which include the guiding-center position ${\bf X}$ and the guiding-center parallel momentum $p_{\|}$, while the gyro-action $J \equiv (mc/e)\,\mu$ (defined in terms of the magnetic moment $\mu$ for a particle of mass $m$ and charge $e$) is canonically-conjugate to the ignorable gyroangle $\theta$. The guiding-center phase-space coordinates $z^{\alpha}$ are constructed by a near-identity transformation from local particle phase-space coordinates $z_{0}^{\alpha} \equiv ({\bf x},p_{\|0},\mu_{0},\theta_{0})$, generated by the phase-space vector fields ${\sf G}_{n}$ 
$(n = 1,2,...)$ \cite{RGL_82}:
\begin{eqnarray}
z^{\alpha} & = & z_{0}^{\alpha} + \epsilon\,G_{1}^{\alpha} + \epsilon^{2} \left( G_{2}^{\alpha} + 
\frac{1}{2}\;{\sf G}_{1}\cdot\exd G_{1}^{\alpha}\right) + \cdots,
\label{eq:gc_trans}
\end{eqnarray}
where the ordering parameter $\epsilon$ appears through the charge renormalization \cite{Northrop_63} $e \rightarrow \epsilon^{-1}e$. 

Hamiltonian guiding-center theory \cite{RGL_83,CB_09} is based on the guiding-center phase-space Lagrangian \cite{RGL_82}
\begin{equation}
\Lambda_{\rm gc} \;\equiv\; {\sf T}_{\rm gc}^{-1}\Lambda_{0} \;+\; \exd S \;\equiv\; \Gamma_{\rm gc} \;-\; H_{\rm gc}\;\exd t,
\label{eq:Lambda_gc}
\end{equation}
where the guiding-center push-forward operator ${\sf T}_{\rm gc}^{-1}$ is generated by $({\sf G}_{1},{\sf G}_{2},\cdots)$ and the gauge-function $S \equiv \epsilon\,S_{1} + \epsilon^{2}S_{2} + \cdots$ is chosen to simplify the form of the guiding-center symplectic one-form $\Gamma_{\rm gc}$. The local phase-space Lagrangian $\Lambda_{0} \equiv \Gamma_{0} - H_{0}\,dt$ in Eq.~\eqref{eq:Lambda_gc} is expressed in terms of the local particle symplectic one-form $\Gamma_{0} \equiv (\epsilon^{-1}e\,{\bf A}/c + p_{\|0}\,\bhat + {\bf p}_{\bot0})\bdot\exd{\bf x}$, with the magnetic field defined as ${\bf B} = \nabla\btimes{\bf A} \equiv B\,\bhat$, and the local particle Hamiltonian $H_{0} \equiv p_{\|0}^{2}/2m + \mu_{0}\,B$. 

The results of the Lie-transform perturbation analysis \eqref{eq:Lambda_gc} are summarized as follows \cite{Brizard_Tronko}. First, the generic form of the guiding-center symplectic one-form is
\begin{eqnarray}
\Gamma_{\rm gc} & \equiv & \left(\frac{e}{c}\,{\bf A} + \Pi_{\|}\,\bhat\right)\bdot\exd{\bf X} \;+\; J\,\left(\exd\theta \;-\frac{}{}{\bf R}\bdot
\exd{\bf X}\right) \nonumber \\
 & \equiv & \frac{e}{c}\,{\bf A}^{*}\bdot\exd{\bf X} \;+\; J\,\exd\theta,
\label{eq:Gamma_gc_def}
\end{eqnarray}
where the vector field ${\bf R}({\bf X})$ ensures that the one-form $\exd\theta - {\bf R}\bdot\exd{\bf X}$ is gyrogauge-invariant \cite{RGL_83}. In 
Eq.~\eqref{eq:Gamma_gc_def}, the symplectic parallel momentum
\begin{equation} 
\Pi_{\|} \;=\; p_{\|} \;+\; \epsilon\,\Pi_{\|}^{(1)} \;+\; \epsilon^{2}\,\Pi_{\|}^{(2)} \;+\; \cdots
\label{eq:Pi_def}
\end{equation}
is defined in terms of the higher-order corrections
\begin{eqnarray}
\Pi_{\|}^{(1)} & = & -\;\left\langle D_{1}\left(p_{\|}\bhat\right)\right\rangle\bdot\bhat \;-\; \frac{1}{2}\left\langle\vb{\rho}_{0}\bdot\nabla\bhat\bdot
{\bf p}_{\bot}\right\rangle, \label{eq:Pi_1_eq} \\
\Pi_{\|}^{(2)} & = & \left\langle D_{2}\left(p_{\|}\bhat\right)\right\rangle\bdot\bhat + \left\langle D_{1}^{2}\left(\frac{p_{\|}}{2}\,\bhat + 
\frac{{\bf p}_{\bot}}{3}\right)\right\rangle\bdot\bhat,
\label{eq:Pi_2_eq}
\end{eqnarray}
where $\langle\cdots\rangle$ denotes gyroangle averaging, $\vb{\rho}_{0} \equiv -\,G_{1}^{\bf x} = (c\bhat/eB)\btimes{\bf p}_{\bot}$ denotes the lowest-order gyroradius, the first-order operator $D_{1}({\bf C}) \equiv {\sf G}_{1}\cdot\exd{\bf C} + \nabla{\bf C}\bdot\vb{\rho}_{0}$ is defined for an arbitrary vector field ${\bf C}$, and we used $D_{n}(p_{\|}\bhat)\bdot\bhat = G_{n}^{p_{\|}} - p_{\|}\,G_{n}^{\bf x}\bdot\vb{\kappa}$ for $n \geq 1$, where $\vb{\kappa} \equiv \bhat\bdot\nabla\bhat$ denotes the magnetic curvature.

Next, the guiding-center Hamiltonian is defined as
\begin{equation}
H_{\rm gc} \;\equiv\; {\sf T}_{\rm gc}^{-1}\,H_{0} \;=\; \frac{p_{\|}^{2}}{2m} \;+\; \Psi, 
\label{eq:Ham_gc_def}
\end{equation}
where the guiding-center ponderomotive potential
\begin{equation}
\Psi \;\equiv\; \mu\,B \;+\; \epsilon\,\Psi_{1} \;+\; \epsilon^{2}\,\Psi_{2} \;+\; \cdots,
\label{eq:Psi_def}
\end{equation}
is defined in terms of the higher-order corrections
\begin{eqnarray}
\Psi_{1} & = & -\; \frac{p_{\|}}{m}\langle G_{1}^{p_{\|}}\rangle \;-\; \langle G_{1}^{\mu}\rangle\;B, \label{eq:Psi_1_eq} \\
\Psi_{2} & = & -\;\mu\;\langle G_{2}^{\bf x}\rangle\bdot\nabla B \;-\; \frac{p_{\|}}{m}\langle G_{2}^{p_{\|}}\rangle \;-\; \langle G_{2}^{\mu}\rangle\;B \nonumber \\
 &  &-\; \frac{1}{2}\;\langle {\sf G}_{1}\rangle\cdot\exd\Psi_{1},
\label{eq:Psi_2_eq}
\end{eqnarray}
where we used $\langle\vb{\rho}_{0}\rangle \equiv 0$ in Eq.~\eqref{eq:Psi_1_eq}. In Eqs.~\eqref{eq:Pi_1_eq}-\eqref{eq:Pi_2_eq} and 
\eqref{eq:Psi_1_eq}-\eqref{eq:Psi_2_eq}, the components $(G_{n}^{\bf x}, G_{n}^{\mu},G_{n}^{\theta})$ are determined from the perpendicular spatial and momentum components of the guiding-center symplectic one-form.

In the perturbation analysis leading to Eqs.~\eqref{eq:Gamma_gc_def} and \eqref{eq:Ham_gc_def}, the generating vector field ${\sf G}_{n}$ is chosen so that the corrections $\Pi_{\|}^{(n)}$ and $\Psi_{n}$ are both independent of the guiding-center gyroangle $\theta$, which yields a guiding-center magnetic-moment adiabatic invariant up to $\epsilon^{n+1}$. Once the guiding-center phase-space Lagrangian \eqref{eq:Lambda_gc} has been constructed to any desired order (we now replace $e/\epsilon$ with $e$), the guiding-center Hamilton equations of motion are expressed in terms of the guiding-center Hamiltonian \eqref{eq:Ham_gc_def} and the guiding-center Poisson bracket constructed from the guiding-center symplectic one-form \eqref{eq:Gamma_gc_def}:
\begin{eqnarray}
\{ F,\; G\}_{\rm gc} & = & \left( \pd{F}{\theta}\,\pd{G}{J} \;-\; \pd{F}{J}\,\pd{G}{\theta} \right) \nonumber \\
 &  &+\; \left( \pd{\Pi_{\|}}{p_{\|}}\right)^{-1}\frac{{\bf B}^{*}}{B_{\|}^{*}}\bdot\left( \nabla^{*}F
\pd{G}{p_{\|}} - \pd{F}{p_{\|}}\nabla^{*}G \right) \nonumber \\
 &  &-\; \frac{c\bhat}{eB_{\|}^{*}}\bdot\left(\nabla^{*}F\btimes\nabla^{*}G\right),
\label{eq:PB_gc_symp}
\end{eqnarray}
where the symplectic magnetic field ${\bf B}^{*} \equiv \nabla\btimes{\bf A}^{*}$ is
\begin{eqnarray}
{\bf B}^{*} & \equiv & {\bf B} \;+\; \frac{c}{e}\; \nabla\btimes\left( \Pi_{\|}\,\bhat \;-\; J\,{\bf R}\right),
\label{eq:B_star}
\end{eqnarray} 
with $B_{\|}^{*} \equiv {\bf B}^{*}\bdot\bhat$, while the gyrogauge-invariant gradient $\nabla^{*} \equiv \nabla + {\bf R}^{*}\,\partial/\partial\theta$ is defined with ${\bf R}^{*} \equiv {\bf R} - \bhat\;\partial\Pi_{\|}/\partial J$. 

The guiding-center equations of motion are now expressed in Hamiltonian form as $d_{\rm gc}z^{\alpha}/dt \equiv \{ z^{\alpha},\; H_{\rm gc}\}_{\rm gc}$, where the guiding-center operator $d_{\rm gc}/dt$ is defined in terms of the particle operator $d/dt$ and the guiding-center push-forward and pull-back operators ${\sf T}_{\rm gc}^{-1}$ and ${\sf T}_{\rm gc}$ generated by $({\sf G}_{1},{\sf G}_{2},\cdots)$ as
\begin{equation}
\frac{d_{\rm gc}}{dt} \;\equiv\; {\sf T}_{\rm gc}^{-1}\;\left(\frac{d}{dt}\;{\sf T}_{\rm gc}\right).
\label{eq:gc_dot_def}
\end{equation}
The reduced guiding-center equations of motion 
\begin{eqnarray}
\frac{d_{\rm gc}{\bf X}}{dt} & = & \left( \pd{\Pi_{\|}}{p_{\|}}\right)^{-1}\pd{H_{\rm gc}}{p_{\|}}\frac{{\bf B}^{*}}{B_{\|}^{*}} + 
\frac{c\,\bhat}{eB_{\|}^{*}}\btimes\nabla H_{\rm gc}, \label{eq:gc_x_symp} \\
\frac{d_{\rm gc}p_{\|}}{dt} & = & -\;\left( \pd{\Pi_{\|}}{p_{\|}}\right)^{-1}\frac{{\bf B}^{*}}{B_{\|}^{*}}\bdot\nabla H_{\rm gc} \label{eq:gc_p_symp}  
\end{eqnarray}
are decoupled from the gyro-motion equations $d_{\rm gc}\theta/dt = \Omega + {\bf R}^{*}\bdot d_{\rm gc}{\bf X}/dt$ and $d_{\rm gc}J/dt = -\;\partial
H_{\rm gc}/\partial\theta \equiv 0$. Moreover, in contrast to non-Hamiltonian guiding-center equations (e.g., \cite{Northrop_63}) the reduced guiding-center Hamilton equations \eqref{eq:gc_x_symp}-\eqref{eq:gc_p_symp} conserve energy ($d_{\rm gc}H_{\rm gc}/dt \equiv 0$) and satisfy the guiding-center Liouville theorem
\begin{equation}
\nabla\bdot\left( {\cal J}_{\rm gc}\;\frac{d_{\rm gc}{\bf X}}{dt} \right) \;+\; \pd{}{p_{\|}}\left( {\cal J}_{\rm gc}\;\frac{d_{\rm gc}p_{\|}}{dt} \right)
\;=\; 0, 
\label{eq:gc_Liouville}
\end{equation}
where ${\cal J}_{\rm gc} \equiv B_{\|}^{*}\;(\partial\Pi_{\|}/\partial p_{\|})$ is the Jacobian for the guiding-center phase-space transformation 
\eqref{eq:gc_trans}. Note that when $\Pi_{\|}$ and $\Psi$ include $n$th-order correction terms, the guiding-center Hamilton equations 
\eqref{eq:gc_x_symp}-\eqref{eq:gc_p_symp} contain terms of order $n+1$ through ${\bf B}^{*}$ and $\nabla\Psi$.

While the reduced guiding-center Hamilton equations \eqref{eq:gc_x_symp}-\eqref{eq:gc_p_symp} are completely general and are valid at all orders in 
$\epsilon$, they can be simplified if the higher-order corrections are placed either entirely in the guiding-center Poisson bracket or in the guiding-center Hamiltonian. In the former case, called the {\it symplectic} representation, the symplectic parallel momentum \eqref{eq:Pi_def} contains all the higher-order corrections (i.e., $\Pi_{\|}^{(n)} \neq 0$ for $n \geq 1$), while the ponderomotive potential \eqref{eq:Psi_def} is simply $\Psi \equiv \mu\,B$ (i.e., $\Psi_{n} \equiv 0$ for $n \geq 1$). In the latter case, called the {\it Hamiltonian} representation, the symplectic parallel momentum $\Pi_{\|}$ replaces $p_{\|}$ as a dynamical variable (i.e., $\Pi_{\|}^{(n)} \equiv 0$ for $n \geq 1$), while the ponderomotive potential 
\eqref{eq:Psi_def} contains the higher-order corrections (i.e., $\Psi_{n} \neq 0$ for $n \geq 1$). 

These two complementary representations are generated by two different phase-space transformations \eqref{eq:gc_trans} leading to the guiding-center coordinates $({\bf X},p_{\|},\mu,\theta)$ generated by (${\sf G}_{1}, {\sf G}_{2}, \cdots$) in the symplectic representation or to the guiding-center coordinates $(\ov{\bf X},\ov{p}_{\|},\ov{\mu},\ov{\theta})$ generated by ($\ov{\sf G}_{1}, \ov{\sf G}_{2}, \cdots$) in the Hamiltonian representation. These representations are said to be equivalent if 
\begin{equation}
(\ov{\bf X},\ov{p}_{\|},\ov{\mu},\ov{\theta}) \;\equiv\; ({\bf X},\Pi_{\|},\mu,\theta),
\label{eq:equivalent}
\end{equation}
so that the Jacobian for the guiding-center transformation leading to the Hamiltonian representation is $\ov{\cal J}_{\rm gc} \equiv B_{\|}^{*}$, since 
$\partial\Pi_{\|}/\partial\ov{p}_{\|} \equiv 1$. The representation equivalence \eqref{eq:equivalent} implies that, for $\alpha \neq p_{\|}$, we have the following relations between generating vector fields
\begin{eqnarray}
\ov{G}_{1}^{\alpha} & \equiv & G_{1}^{\alpha}, \label{eq:alpha_1} \\
\ov{G}_{2}^{\alpha} & \equiv & G_{2}^{\alpha} \;-\; \frac{1}{2}\,\Pi_{\|}^{(1)}\;\pd{G_{1}^{\alpha}}{p_{\|}},
\label{eq:alpha_2}
\end{eqnarray}
while, using Eq.~\eqref{eq:Pi_def}, we have
\begin{eqnarray}
\ov{G}_{1}^{p_{\|}} & \equiv & G_{1}^{p_{\|}} \;+\; \Pi_{\|}^{(1)}, \label{eq:p_1} \\
\ov{G}_{2}^{p_{\|}} & \equiv & G_{2}^{p_{\|}} + \Pi_{\|}^{(2)} + \frac{1}{2}\left({\sf G}_{1}\cdot\exd G_{1}^{p_{\|}} - \ov{\sf G}_{1}\cdot\exd 
\ov{G}_{1}^{p_{\|}}\right).
\label{eq:p_2}
\end{eqnarray} 
The representation-invariance of the magnetic moment $\ov{\mu} \equiv \mu$ implies that, once the corrections $\Pi_{\|}^{(n)}$ are found in the symplectic representation, the corrections $\Psi_{n}$ are immediately found in the Hamiltonian representation as
\begin{eqnarray}
\Psi_{1} & = & -\;\frac{p_{\|}}{m}\;\Pi_{\|}^{(1)}, \label{eq:psi1_equivalence} \\
\Psi_{2} & = & -\;\frac{p_{\|}}{m}\;\Pi_{\|}^{(2)} \;+\; \frac{1}{2m}\;\left(\Pi_{\|}^{(1)}\right)^{2}.
\label{eq:psi2_equivalence}
\end{eqnarray}
Hence, once a higher-order Hamiltonian guiding-center theory is derived in one representation, one can also construct a theory in the complementary representation. More importantly, it can also be shown \cite{Brizard_Tronko} that the guiding-center Hamilton equations 
\eqref{eq:gc_x_symp}-\eqref{eq:gc_p_symp} are identical in all equivalent representations.

We now proceed with the construction of a higher-order Hamiltonian guiding-center theory that retains up to second-order corrections. While it appears that the derivation of the $n$th-order corrections $\Pi_{\|}^{(n)}$ and $\Psi_{n}$ requires obtaining the guiding-center phase-space transformation 
\eqref{eq:gc_trans} up to $n$th order, we now show that, by introducing the perturbative guiding-center Ba\~{n}os hierarchy in the symplectic representation, we obtain the remarkable fact that the $n$th-order Hamiltonian structure can be determined from the guiding-center phase-space transformation \eqref{eq:gc_trans} up to $(n-1)$th order only.

The guiding-center Ba\~{n}os hierarchy is constructed in the symplectic representation as follows. Using the functional definition \eqref{eq:gc_dot_def} for $d_{\rm gc}/dt$, we introduce the identity
\begin{eqnarray}
{\sf T}_{\rm gc}^{-1}\left(\frac{d{\bf x}}{dt}\right) & \equiv & \frac{d_{\rm gc}}{dt}\left({\sf T}_{\rm gc}^{-1}{\bf x} \right) = 
\frac{d_{\rm gc}{\bf X}}{dt} + \frac{d_{\rm gc}\vb{\rho}_{\rm gc}}{dt}
\label{eq:gc_id} 
\end{eqnarray}
expressed in terms of the guiding-center velocity $d_{\rm gc}{\bf X}/dt$ and the guiding-center displacement velocity $d_{\rm gc}\vb{\rho}_{\rm gc}/dt$ (which includes the polarization velocity $d\langle\vb{\rho}_{\rm gc}\rangle/dt$). Here, the guiding-center displacement is expanded as
\begin{equation}
\vb{\rho}_{\rm gc} \;\equiv\; {\sf T}_{\rm gc}^{-1}{\bf x} \;-\; {\bf X} \;=\; \epsilon\,\vb{\rho}_{0} \;+\; \epsilon^{2}\;
\vb{\rho}_{1} \;+\; \epsilon^{3}\;\vb{\rho}_{2} \;+\; \cdots,
\label{eq:rho_gc}
\end{equation}
where the higher-order corrections
\begin{eqnarray}
\vb{\rho}_{1} & = & -\;G_{2}^{\bf x} \;-\; \frac{1}{2}\;{\sf G}_{1}\cdot\exd\vb{\rho}_{0}, \label{eq:rho_1} \\
\vb{\rho}_{2} & = & -\;G_{3}^{\bf x} - {\sf G}_{2}\cdot\exd\vb{\rho}_{0} + \frac{1}{6}\,{\sf G}_{1}\cdot
\exd({\sf G}_{1}\cdot\exd\vb{\rho}_{0}) \label{eq:rho_2}
\end{eqnarray}
satisfy $\langle\vb{\rho}_{n}\rangle \neq 0$ and $\vb{\rho}_{n}\bdot\bhat \neq 0$. By taking the dot product of Eq.~\eqref{eq:gc_id} with $m\,\bhat$ (defined in guiding-center phase space), we obtain the guiding-center Ba\~{n}os hierarchy
\begin{equation}
{\sf T}_{\rm gc}^{-1}\left[\left({\sf T}_{\rm gc}\frac{}{}\bhat\right)\bdot m\,\frac{d{\bf x}}{dt}\right] \;=\; m\,\bhat\bdot\left(
\frac{d_{\rm gc}{\bf X}}{dt} \;+\; \frac{d_{\rm gc}\vb{\rho}_{\rm gc}}{dt}\right),
\label{eq:Banos_hierarchy}
\end{equation}
where we used the identity $f\;({\sf T}_{\rm gc}^{-1}g) \equiv {\sf T}_{\rm gc}^{-1}({\sf T}_{\rm gc}f\;g)$. The left side of the guiding-center 
Ba\~{n}os hierarchy \eqref{eq:Banos_hierarchy} is expanded in powers of $\epsilon$ as
\begin{eqnarray}
{\sf T}_{\rm gc}^{-1}\left[\left({\sf T}_{\rm gc}\frac{}{}\bhat\right)\bdot m\,\frac{d{\bf x}}{dt}\right] & = & \left( p_{\|} \;-\; \epsilon\;
G_{1}^{p_{\|}} \;+\; \cdots \right) \nonumber \\
 &  &-\, \epsilon\,\vb{\rho}_{0}\bdot\nabla\bhat\bdot{\bf p}_{\bot} + \cdots,
\label{eq:Banos_drift}
\end{eqnarray}
where the gyroangle-averaged first-order correction
\begin{equation}
-\;\left\langle\vb{\rho}_{0}\bdot\nabla\bhat\bdot{\bf p}_{\bot}\right\rangle \;=\; J\tau
\label{eq:Banos_first}
\end{equation}
represents the Ba\~{n}os parallel momentum \cite{Banos,Northrop_Rome}, defined in terms of the {\it twist} of the magnetic-field lines $\tau \equiv \bhat\bdot\nabla\btimes\bhat$. The right side of the guiding-center Ba\~{n}os hierarchy \eqref{eq:Banos_hierarchy} involves 
\begin{eqnarray}
m\;\bhat\bdot\frac{d_{\rm gc}{\bf X}}{dt} & = &  p_{\|} \;-\; \epsilon\;p_{\|}\,\pd{\Pi_{\|}^{(1)}}{p_{\|}} \;+\; \cdots, \label{eq:Banos_X} \\
m\bhat\bdot\frac{d_{\rm gc}\vb{\rho}_{\rm gc}}{dt} & = & \epsilon\;m\bhat\bdot\left(\Omega\,\pd{\vb{\rho}_{1}}{\theta} + 
\frac{d_{0}\vb{\rho}_{0}}{dt} \right) + \cdots, \label{eq:Banos_rho}
\end{eqnarray}
where $\bhat\bdot\partial\vb{\rho}_{0}/\partial\theta \equiv 0$ and the guiding-center operator \eqref{eq:gc_dot_def} is expanded as
\begin{equation}
\frac{d_{\rm gc}}{dt} \;\equiv\; \epsilon^{-1}\Omega\;\pd{}{\theta} \;+\; \frac{d_{0}}{dt} \;+\; \epsilon\;\frac{d_{1}}{dt} \;+\;
\cdots.
\label{eq:dgc_dt}
\end{equation}
The guiding-center Ba\~{n}os hierarchy \eqref{eq:Banos_hierarchy} is a perturbative hierarchy that relates $\langle G_{n}^{p_{\|}}\rangle$ with $\partial\Pi_{\|}^{(n)}/\partial p_{\|}$ at each order $n \geq 1$. At first order, the guiding-center Ba\~{n}os hierarchy \eqref{eq:Banos_hierarchy} yields the gyroangle-averaged relation
\begin{equation}
\langle G_{1}^{p_{\|}}\rangle \;=\; p_{\|}\;\pd{\Pi_{\|}^{(1)}}{p_{\|}} \;+\; J\,\tau.
\label{eq:Banos_G1_av}
\end{equation}
When it is combined with Eq.~\eqref{eq:Pi_1_eq}, with Eq.~\eqref{eq:Banos_first}:
\begin{equation}
\langle G_{1}^{p_{\|}}\rangle \;=\; -\,\Pi_{\|}^{(1)} \;+\; \frac{1}{2}\;J\,\tau 
\end{equation}
we obtain the first-order Ba\~{n}os equation
\begin{equation}
\pd{}{p_{\|}}\left(p_{\|}\frac{}{}\Pi_{\|}^{(1)} \right) \;=\; -\;\frac{1}{2}\;J\,\tau,
\label{eq:Pi_1_Banos}
\end{equation}
where the unknown $\langle G_{1}^{p_{\|}}\rangle$ has been eliminated. Integrating Eq.~\eqref{eq:Pi_1_Banos} with respect to $p_{\|}$ yields the first-order correction
\begin{equation}
\Pi_{\|}^{(1)} \;=\; -\;\frac{1}{2}\;J\,\tau.
\label{eq:Pi_1}
\end{equation}
This solution means that, since $\langle G_{1}^{p_{\|}}\rangle = J\,\tau$ in the symplectic representation, the first-order Ba\~{n}os parallel momentum 
$J\,\tau$ is hidden in the definition of the guiding-center parallel momentum. Note that the first-order symplectic momentum \eqref{eq:Pi_1} was obtained without knowing the details of the first-order perturbation analysis \cite{RGL_83,CB_09}.

At second order, we similarly obtain the second-order Ba\~{n}os equation
\begin{equation}
\pd{}{p_{\|}}\left( p_{\|}\frac{}{} \Pi_{\|}^{(2)}\right) \;\equiv\; 2\,p_{\|}\,\varrho_{\|}^{2}\,|\vb{\kappa}|^{2} \;-\;
J\,\varrho_{\|}\,\beta_{2},
\label{eq:Pi2_ave_final}
\end{equation}
where $\varrho_{\|} \equiv p_{\|}/(m\Omega)$ and the scalar field
\begin{eqnarray}
\beta_{2} & \equiv & 3\,\vb{\kappa}\bdot\left(\vb{\kappa} \;-\frac{}{} \nabla\ln B \right) \;-\; \frac{1}{4}\left[ \tau^{2} \;+\; (\nabla\bdot\bhat)^{2} \right] \nonumber \\
 &  &+\; \frac{1}{2}\;\nabla\bdot\left[\vb{\kappa} \;+\frac{}{} \bhat\;(\nabla\bdot\bhat)\right]
\label{eq:beta_2}
\end{eqnarray}
involves terms of second order in magnetic-field nonuniformity. Integrating Eq.~\eqref{eq:Pi2_ave_final} with respect to $p_{\|}$ yields the second-order correction
\begin{equation}
\Pi_{\|}^{(2)} \;=\; \frac{1}{2}\,\left( p_{\|}\;\varrho_{\|}^{2}\,|\vb{\kappa}|^{2} \;-\frac{}{} J\,\varrho_{\|}\;\beta_{2} \right),
\label{eq:Pi2_ave_sol}
\end{equation}
which is obtained solely from the details of the first-order perturbation analysis \cite{RGL_83,CB_09}.

In the symplectic representation (with $\Psi = \mu\,B$), the guiding-center symplectic parallel momentum is thus
\begin{eqnarray}
\Pi_{\|} & \equiv & p_{\|}\left( 1 + \frac{\varrho_{\|}^{2}}{2}\,|\vb{\kappa}|^{2} \right) - \frac{J}{2}\,\left( \tau +\frac{}{} 
\varrho_{\|}\,\beta_{2} \right) + \cdots,
\label{eq:Pi||_sol}
\end{eqnarray}
so that, up to second order, the guiding-center symplectic one-form is
\begin{eqnarray}
\Gamma_{\rm gc} & \equiv & \left[ \frac{e}{c}\,{\bf A} \;+\; p_{\|}\; \left( 1 \;+\; \frac{1}{2}\;
\varrho_{\|}^{2}\,|\vb{\kappa}|^{2}\right)\;\bhat \right]\bdot\exd{\bf X} \nonumber \\
 &  &+\; J\,\left(\exd\theta \;-\; {\bf R}^{*}\bdot\exd{\bf X}\right),
\label{eq:Gamma_gc_symp}
\end{eqnarray}
where $\varrho_{\|} \equiv \Pi_{\|}/(m\Omega)$ and ${\bf R}^{*} \equiv {\bf R} + \frac{1}{2}\;( \tau + \varrho_{\|}\,\beta_{2})\;\bhat$, while
Eq.~\eqref{eq:gc_x_symp} involves the factor
\begin{equation}
\pd{\Pi_{\|}}{p_{\|}} \;=\; 1 \;+\; \frac{3}{2}\,\varrho_{\|}^{2}\,|\vb{\kappa}|^{2} \;-\; \frac{J\,\beta_{2}}{2\,m\Omega},
\end{equation}
where the first-order corrections vanish.

In the Hamiltonian representation (where $\Pi_{\|}$ acts as a dynamical variable), on the other hand, the equivalence relations \eqref{eq:psi1_equivalence}-\eqref{eq:psi2_equivalence} are used to construct the guiding-center ponderomotive potential
\begin{eqnarray}
\Psi & = & \mu\,B\left( 1 \;+\; \frac{1}{2}\,\varrho_{\|}\tau \;+\; \frac{1}{8}\;\frac{J\,\tau^{2}}{m\Omega} \right) \nonumber \\
 &  &+\; \frac{\Pi_{\|}^{2}}{2m}\; \left( \frac{J\,\beta_{2}}{m\Omega} \;-\; \varrho_{\|}^{2}\,|\vb{\kappa}|^{2}\right).
\label{eq:Psi_Ham}
\end{eqnarray}
All second-order terms in Eq.~\eqref{eq:Psi_Ham} appear in Eq.~(135) of Parra and Calvo \cite{Parra_Calvo}. However, since they use a mixed representation (i.e., $\Pi_{\|}^{(1)} \neq 0$ and $\Psi_{2} \neq 0$), which may not be equivalent to ours, there are additional terms as well (which do not satisfy the Ba\~{n}os hierarchy).

In summary, we have presented two equivalent representations of higher-order Hamiltonian guiding-center theory that retains second-order corrections and satisfies the perturbative Ba\~{n}os hierarchy \eqref{eq:Banos_hierarchy} derived from the guiding-center identity \eqref{eq:gc_id}. These representations yield identical Hamilton guiding-center equations, which can be used to construct a consistent Hamiltonian gyrocenter theory used to study the long-time magnetic confinement of charged particles in the presence of electromagnetic fluctuations.

Work by AJB was supported by U.~S.~DoE grant No.~DE-FG02-09ER55005, while work by NT was supported by CNRS (ANR EGYPT) and Euratom-CEA (contract EUR 344-88-1 FUA F) and  EPSRC S \& I grant No.~EP/D062837.


\begin{thebibliography}{99}
\bibitem{Brizard_Tronko} Details of the Lie-transform perturbation analysis will appear in a forthcoming publication.
\end{thebibliography}
\end{document}